\documentstyle[epsf]{mn}

\def\delchi{{$\Delta\chi^{2}$}}
\def\etal{{et al.}}

\def\asca{{\it ASCA}}

\def\chisq{{$\chi^{2}$}}

\def\eg{{e.g.,\ }}
\def\Msun{\hbox{$\rm\thinspace M_{\odot}$}}

\def\mnras{{MNRAS}}

\def\nat{Nat}
\def\apj{{ApJ}}

\def\apj{{ApJ}}

\title {VARIABILITY OF THE IRON K-EMISSION LINE IN THE SEYFERT 1
GALAXY NGC 3516}

\author[K. Nandra et al.]{K. Nandra$^{1,2}$ 
	R.F. Mushotzky$^{1}$, 
	T. Yaqoob$^{1,3}$, 
	I.M. George$^{1,3}$,
	T.J. Turner$^{1,3}$  \\
\small
$^{1}$Laboratory for High Energy Astrophysics, Code 660,
	NASA/Goddard Space Flight Center,
  	Greenbelt, MD 20771 \\
$^{2}$NAS/NRC Research Associate \\
$^{3}$Universities Space Research Association \\
}

\date{To appear in Mon. Not. R. Astron. Soc.}

\begin{document}

\maketitle

\begin{abstract}

We present strong evidence for variability of the flux of the iron
K$\alpha$ emission line in the Seyfert 1 galaxy, NGC 3516. 
Two ASCA observations, 
separated by $\sim 1$~yr, showed a marked decrease in 
continuum flux by $\sim 60$~per cent. The flux in the broad,
iron K$\alpha$ line decreased by the same factor in this time period,
with no evidence for changes in the line profile.
The line variability is significant at $>99$~per cent confidence
and rules out models in which the line is produced in a molecular
torus located at $>1$~pc from the nucleus. An accretion disk is
considerably more likely. 

\end{abstract}


\section{INTRODUCTION}
\label{Sec:Introduction}

Recent \asca\ observations have shown that the iron lines commonly
observed in Seyfert 1 galaxies (Nandra \& Pounds 1994) are extremely
broad, with FWHM of order $50,0000$~km s$^{-1}$ (Mushotzky \etal\
1995; Tanaka \etal\ 1995; Yaqoob \etal\ 1995; Nandra \etal\ 1996b,
hereafter N96).  These observations alone provide strong evidence for
a black hole/accretion disk system, the line widths being extremely
difficult to account for in any other geometry (Fabian \etal\ 1995). 
A specific prediction of the disk-line model is that the line should
respond rapidly to changes in the continuum, as it is excited by
fluorescence in a region close to the central black hole. Evidence for
such line variability in Seyfert 1 galaxies has proved surprisingly
elusive. The well-studied case of the bright Seyfert galaxy NGC 4151
failed to provide conclusive evidence, despite over 10 years worth of
high-quality data (\eg Warwick \etal\ 1989).  The best cases reported
to date have been those of MCG-6-30-15 and NGC 7314, for which Iwasawa
\etal\ (1996) and Yaqoob \etal\ (1996) have shown changes in the iron
K$\alpha$ profile. Interestingly, in the former case, the variations
were such that the narrow and broad components of the line were
anti-correlated, meaning that the evidence for variability of the
total flux of the line was weak. This is contrary to the predictions
of simple accretion disk models, where the line should track the
continuum variations in a linear fashion, even on short time scales.
Thus it is clearly essential to examine the variations in other
sources, to investigate whether or not such behaviour is common.

Here we present two ASCA observations of the bright Seyfert 1
galaxy NGC 3516, separated by a 1 year baseline. This source has
shown strong continuum variability in the past, allowing us to search
for the associated variations expected in the emission line.

\section{OBSERVATIONS}

NGC 3516 was observed by \asca\ (Tanaka, Inoue \& Holt 1994)
on 1994-Apr-02 (hereafter observation 1) and 1995-March-12 
(hereafter observation 2) using both Solid--state Imaging
Spectrometers (SIS0/SIS1), and the two Gas Imaging
Spectrometers (GIS2/GIS3).

The SIS data were collected in 1-CCD readout mode using {\tt FAINT}
telemetry mode. The data were analyzed according to the methods
described in Nandra \etal\ (1996a) and N96. The reader is referred to
references within those papers for further details regarding ASCA
analysis.

Source events were extracted for all four detectors using a circular
cell, centred on NGC 3516, with typical radii 3-4 arcmin. The
background was estimated using source-free regions. The total exposure
times after screening were of order 30ks for observation 1 and $40$~ks
for observation 2. Examination of the total count rates for the two
observations immediately suggests flux variability. Using similar
extraction cells, we obtained $2.50\pm 0.01$ ct s$^{-1}$ (0.4-10 keV)
in SIS0 for observation 1 and $1.52\pm 0.01$ ct s$^{-1}$ for
observation 2, in the same instrument. We proceeded immediately to
analysis of the spectra.

\section{LINE PROPERTIES}

The X-ray spectrum of NGC 3516 is absorbed by a column of partially
ionized gas (e.g., Kriss \etal\ 1996). However, this warm absorber has
a negligible effect on the spectrum above $\sim 3$~keV in observation
1 (George \etal\ 1996). We have verified that this is also the case
for observation 2 and to explore the properties of the line have used
only the data in the 3-10 keV band.

As shown by N96, NGC 3516 exhibits a line profile typical of Seyfert 1
galaxies, which is characteristic of an accretion disk orbiting a
central black hole (e.g., Fabian \etal\ 1989). We show in Fig.~1 the
line profiles determined from the two observations by fitting a
power-law model to the SIS data in the 3-5 and 7-10 keV ranges, i.e.
excluding the ``iron band''. 
\begin{figure}

\centering 
\epsfxsize=0.5\textwidth
\epsffile{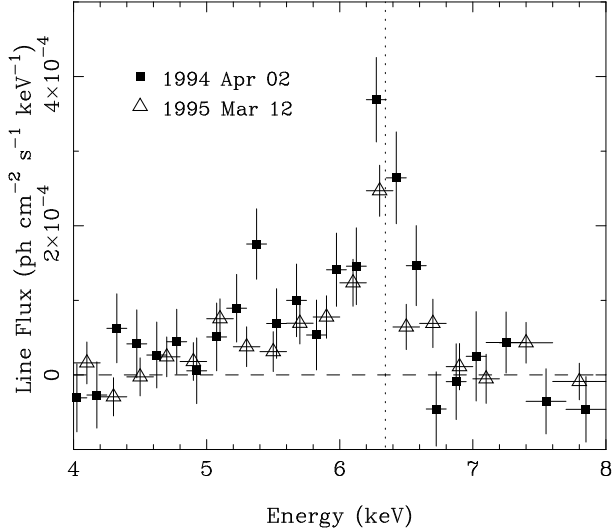}

\caption
{Combined SIS0/SIS1 line profiles for the two observations, which have
been created by interpolating a power law continuum excluding the 5-7
keV region.  The solid squares represent the data for observation 1,
the open triangles those for observation 2. The line flux changes
between the two observations, but there is no obvious, or
statistically significant, change in profile.}
\end{figure}
Several things are apparent from this
figure.  First, the iron K$\alpha$ line is very broad and asymmetric
in this object, as has already been shown by N96. Furthermore, the
profiles determined for the two observations are similar. Most
interestingly from the point-of-view of the present paper, however, is
the clear change in line flux, particularly evident in the core
(6.1-6.7~keV).  To quantify this further and estimate the statistical
significance of this change in line flux, we have employed the
disk-line model of Fabian \etal\ (1989) in addition to the power law
in the 3-10 keV band.  Such a model provides an excellent fit to both
datasets. However, there are too many free parameters in that model to
provide a clear indication of the line variability. We therefore made
the assumption, justified by Fig.~1, that the line profile remained
constant between the two observations, and adopted the disk-line
parameters determined by N96 to define that profile. These are a rest
energy of $6.4$~keV, an inner radius of $6\ R_{\rm g}$ (where $R_{\rm
g}$ is the gravitational radius of the black hole), an outer radius of
$1000$~$R_{\rm g}$, a disk inclination of $27^{\circ}$ and an emissivity
profile varying as $R^{-q}$ with $q=2.7$.  These parameters are fairly
typical for Seyfert galaxies in general. We then fitted the 3-10 keV
spectra allowing the power-law parameters and the normalization of the
emission line to vary.

The confidence contours obtained from those fits are shown in Fig.~2.
\begin{figure}

\centering 
\epsfxsize=0.5\textwidth
\epsffile{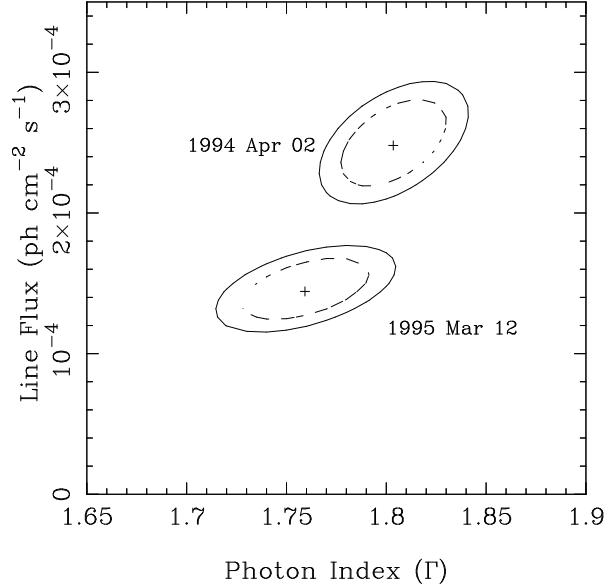}

\caption
{Confidence contours for the photon index ($\Gamma$) and line
intensity obtained from a fit with a model consisting of a power law
plus disk line. The contours are 68 (dashed lines) and 90 per cent
(solid lines) for two interesting parameters. The line decreases in
flux in proportion to the continuum, at $>99$~per cent
confidence. There is no statistically significant change in the
continuum slope.}
\end{figure} 
This clearly shows that, while the continuum slope is consistent
between the two observations, the line flux varied significantly. The
90~per cent confidence regions for the line do not overlap, indicating
variability of the line at $>99$~per cent confidence. The spectral
indices are $\Gamma=1.80^{+0.03}_{-0.02}$ for observation 1 and
$1.76^{+0.03}_{-0.03}$ for observation 2 (68 per cent confidence for
two interesting parameters). The line fluxes are $2.5^{+0.3}_{-0.3}$
and $1.5^{+0.2}_{-0.3}$, both in units of $10^{-4}$~ph cm$^{-2}$
s$^{-1}$. The equivalent widths are remarkably consistent at
$330^{+40}_{-40}$~eV and $320^{+40}_{-60}$~eV respectively. This
suggests that the line varied in strict proportion with the
continuum. The continuum flux reduced by a factor 1.7 between the two
observations and the line flux by a factor $1.7\pm 0.4$. 

If the line arises by X-ray illumination of an accretion disk, we
expect it to be accompanied by a ``reflection'' component due to
Compton scattering in the disk (\eg\ George \& Fabian 1991; Matt
\etal\ 1991). The presence of such a component does not affect our
conclusions regarding the line variability. Including a reflection
component appropriate for a disk inclined at $27\deg$, with a covering
fraction of $2\pi$ reduces both line fluxes by $\sim 20$~per cent to
$2.1\pm 0.3$ and $1.2\pm 0.2$ $10^{-4}$~ph cm$^{-2}$ s$^{-1}$
respectively.  Clearly the change is still highly significant.
Note, however, that the best-fit power-law index is steeper,
with $\Gamma\sim 1.9$, when reflection is included. These effects have
already been noted by N96.

Finally, we have investigated whether small variations in the line
shape affect our results. Although no large changes are evident in
Fig.~1, we have allowed for this by allowing $q$ to be free in the fit
to each dataset. Fig.~3 shows the change in the confidence surfaces
when we allow the line profile to vary. 
\begin{figure}
\centering 
\epsfxsize=0.5\textwidth
\epsffile{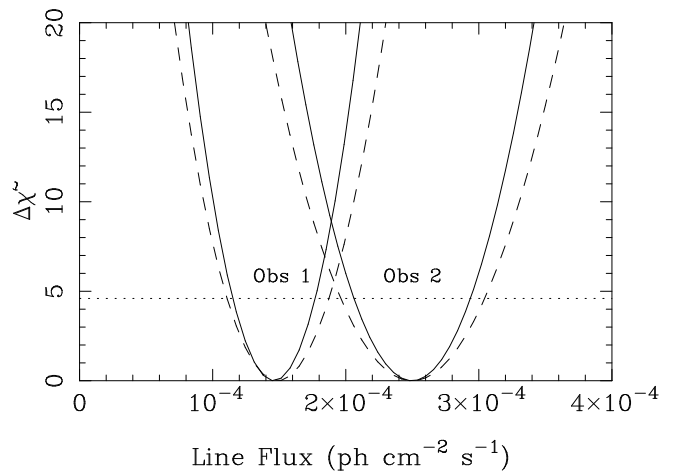}
\caption
{Change in \chisq\ when the line flux is varied from its
best-fit value. The solid lines represent the confidence surface when
$q$ is fixed (i.e. a projection of Fig.~2 into one dimension). The
dashed lines show the effect of allowing $q$ to be variable. There is
little change.  The horizontal dotted line represents \delchi=4.6, 90
per cent confidence for two interesting parameters. The fact that the
contours do not overlap below this line implies variability of the
line at $>99$~per cent confidence.  For the case where $q$ is free,
we can consider 3 parameters to be interesting ($q$, line flux and photon
index). Even in this most conservative case, where we allow the line
profile to change, we still infer that the line flux is variable, in
this instance at $>95$~~per cent confidence}
\end{figure}
Clearly the parameter values
are not as strongly constrained as when $q$ was fixed, however, we
still conclude that the line is variable.

\section{DISCUSSION}

We have presented evidence for variability of the iron K$\alpha$
emission line in NGC 3516. The line flux changes proportionately with
the continuum over a $\sim 1$~yr baseline. This sets a firm upper
limit of 1 lt-yr ($R_{\rm max} \sim 10^{18}$~cm) for the extent of the
line-producing region. It is likely to be much smaller than this; the
duration of the individual observations is $\sim 1$~d and if the
fluorescing material extended over a region much larger than this, the
fact that the line changes by the same factor as the continuum would
have to be considered as co-incidental, as it would be responding to
the continuum averaged over the previous $R_{\rm max}/c$.  As NGC 3516
shows strong short time scale (i.e. intra-day) variability (Kolman
\etal\ 1993; Nandra \etal\ 1996a), we therefore conclude that the line
is most likely produced in a region $<1$~lt day in extent. This agrees
very well with the predictions of accretion disk models, which are
required to explain the broad, asymmetric profiles (Tanaka \etal\
1995; N96). Assuming for the moment that $R_{\rm max}<1$~lt day), this
corresponds to a distance of $\sim 200/M_{8}\ R_{\rm g}$ gravitational
radii, where $M_{8}$ is the black hole mass in units of
$10^{8}$~\Msun. The average line properties for Seyfert 1 galaxies
suggest that around $80$~per cent of the line emission arises within
$\sim 100\ R_{\rm g}$ of the nucleus.

The fact that the K$\alpha$ line is consistent with a linear change
with flux suggests that it originates predominantly from one region.
Indeed, the clear change in the core is strongly indicative that
this flux comes from close to the central engine. An alternative
hypothesis, that the broad component arises from an accretion disk but
that the line core is produced in a molecular torus, located far from
the nucleus (Ghisellini, Haardt \& Matt 1994; 
Krolik, Madau \& Zycki 1994) is strongly
disfavored. We conclude that the bulk of the iron K$\alpha$
emission in NGC 3516, and most likely in the majority of Seyfert 1
galaxies, arises in an accretion disk extremely close to the central
black hole.

It is interesting to compare our results with those presented for
MCG-6-30-15 by Iwasawa et al. (1996). In that source, while the narrow
core of the line was found to be well correlated with the continuum
variations, the broad wing was anti correlated, at least on medium
time scales ($\sim$1 day). This is somewhat contrary to the behaviour
expected in simple accretion disk models, and may require the
invocation of relativistic effects within the X-ray source itself
and/or changes in the pattern of X-ray illumination, as well as the
light bending and gravitational effects characteristic of a Kerr black
hole. On the other hand NGC 7314 (Yaqoob et al. 1996) shows more
variability in the wing than the core, as predicted. This may also be
true for MCG-6-30-15 on the shortest time scales (Iwasawa et
al. 1996). Our results show no apparent change in profile, which is to
be expected over the longer observation baseline (1~yr), where the
total flux of the line should track the continuum. The stability of
the line profile over this time period suggests that there are no
gross changes in the geometry of the system (accretion disk and X-ray
source). Further progress requires better-sampled data with high
signal-to-noise-ratio, but the results so far suggest that, even with
current instrumentation, such observations would be most rewarding.

\section*{ACKNOWLEDGMENTS}

We thank the \asca\ team for their operation of the satellite, and the
\asca\ GOF at NASA/GSFC for helpful discussions. We acknowledge the
financial support of the National Research Council (KN) and
Universities Space Research Association (IMG,TJT). This research has
made use of data obtained through the HEASARC on-line service, provided
by NASA/GSFC.


\section*{REFERENCES}

\def\reference{\par\parskip 0pt\noindent\hangindent 20pt}

\reference Fabian, A.C., Nandra, K., Reynolds, C.S., Brandt, W.N. Otani, C.,
	   Tanaka, Y.,
	1995, \mnras, 277, L11
\reference Fabian, A.C., Rees, M.J., Stella, L., White, N.E.,
	1989, \mnras, 238, 729
\reference George, I.M., Fabian, A.C.,
	1991, \mnras, 249, 352
\reference George, I.M., \etal, 
	1996, in preparation
\reference Ghisellini, G., Haardt, F., Matt, G., 
	1994, \mnras, 267, 743
\reference Iwasawa, K., \etal,
	1996, \mnras, in press
\reference Kolman, M., Halpern, J.P., Martin, C., Awaki, H., Koyama, K.,
	1993, ApJ, 403, 592
\reference Kriss, G.A., \etal,
	1996, \apj, in press
\reference Krolik, J.H., Madau, P., Zycki, P.T.,
	1994, \apj, 420, L57
\reference Matt, G., Perola, G.C., Piro, L., 
        1991, A\&A, 245, 63
\reference Mushotzky, R.F., Fabian, A.C., Iwasawa, K., Matsuoka, M., 
           Nandra, K., Tanaka, Y., 
	1995, \mnras, 272, L9
\reference Nandra, K., Pounds, K.A.,
	1994, \mnras, 268, 405
\reference Nandra, K., George, I.M., Mushotzky, R.F., Turner, T.J., Yaqoob, T.,
	1996a, \apj, in press
\reference Nandra, K., George, I.M., Mushotzky, R.F., Turner, T.J., 
                 Yaqoob, T.,
	1996b, \apj, in press (N96)
\reference Tanaka, Y., \etal,
	1995, \nat, 375, 659
\reference Tanaka, Y., Inoue, H., Holt, S.S.,
	1994, PASJ, 46, L37
\reference Warwick, R.S., Yaqoob, T., Pounds, K.A., Matsuoka, M., 
                 Yamauchi, M.,
	1989, PASJ, 41, 721
\reference Yaqoob, T., Edelson, R., Weaver, K.A., Warwick, R.S., 
	   Mushotzky, R.F., Serlemitsos, P.J., Holt, S.S.,
	1995, \apj, 453, L81
\reference Yaqoob, T., Serlemitsos, P.J., Turner, T.J., George, I.M.,
	   Nandra, K.,
	1996, \apj, submitted
\clearpage

\end{document}